\documentclass[letterpaper,11pt]{article}

\usepackage{scrextend}
\usepackage{amsmath,amssymb,epsfig,bbm}
\usepackage[active]{srcltx}
\usepackage{amsrefs}
\usepackage{MnSymbol}
\usepackage{slashed}
\usepackage{amsthm}

\usepackage{bbm}

\usepackage{dsfont}

\newcommand{\be}{\begin{equation}}
\newcommand{\ee}{\end{equation}} 
\newcommand{\ba}{\begin{eqnarray}}
\newcommand{\ea}{\end{eqnarray}}
\newcommand{\nn}{\nonumber}

\def\d{\delta}
\def\e{\epsilon}

\def\m{\mu}
\def\n{\nu}

\def\s{\sigma}

\def\ca{{\cal A}}
\def\cb{{\cal B}}

\def\cf{{\cal F}}

\def\R{{\mathbb{R}}}

\makeatletter
\newcommand{\eqnum}{\refstepcounter{equation}\textup{\tagform@{\theequation}}}
\makeatother

\newcommand{\C}{{\Bbb C}}

\newtheorem{thm}{Theorem}[subsection]

\newtheorem{prop}[thm]{Proposition}

\newtheorem{definition}[thm]{Definition}

\newtheorem*{definition*}{Definition}

\newcommand{\g}{{\frak{g}}}

\fontfamily{yfrak}

\begin{document}

\vskip 25mm

\begin{center}

{\large\bfseries  
Representations of the Quantum\\ Holonomy-Diffeomorphism Algebra
}

\vskip 6ex

Johannes \textsc{Aastrup}$^{a}$\footnote{email: \texttt{aastrup@math.uni-hannover.de}} \&
Jesper M\o ller \textsc{Grimstrup}$^{b}$\footnote{email: \texttt{jesper.grimstrup@gmail.com}}\\ 
\vskip 3ex

$^{a}\,$\textit{Institut f\"ur Analysis, Universit\"at Hannover, \\ Welfengarten 1, 
D-30167 Hannover, Germany.}
\\[3ex]
$^{b}\,$\textit{Independent researcher, Copenhagen, Denmark.
}\\[3ex]

{\footnotesize\it This work is financially supported by Ilyas Khan, \\\vspace{-0.1cm}St EdmundÕs College, Cambridge, United Kingdom.}

\end{center}

\vskip 3ex

\begin{abstract}
In this paper we continue the development of Quantum Holonomy Theory, which is a candidate for a fundamental theory, by constructing separable strongly continuous representations of its algebraic foundation, the quantum holonomy-diffeomorphism algebra. Since the quantum holonomy-diffeomorphism algebra encodes the canonical commutation relations of a gauge theory these representations provide a possible framework for the kinematical sector of a quantum gauge theory. Furthermore, we device a method of constructing physically interesting operators such as the Yang-Mills Hamilton operator. 
This establishes the existence of a general non-perturbative framework of quantum gauge theories on a curved backgrounds. Questions concerning gauge-invariance are left open.

\end{abstract}

\newpage
\tableofcontents

\section{Introduction}

One of the key unsolved problems in contemporary theoretical physics is the rigorous understanding of quantum field theory and in particular quantum gauge theory. In this paper, which is a part of a larger research project called Quantum Holonomy Theory \cite{A3}, we show that Hilbert space representation of an elementary algebra based on holonomy-diffeomorphisms exist. This algebra, which we call the $\mathbf{QHD}(M)$ algebra \cite{MR3482818}, encodes the canonical commutation relations of a quantum gauge theory and therefore constitutes a natural foundation for a non-perturbative framework for a quantum gauge theory.\\

To understand the $\mathbf{QHD}(M)$ algebra let us first consider quantum mechanics.
The usual way to construct quantum mechanics on for example the real line $\R$ is to consider $L^2(\R )$ together with the operators $x$ and $  i \tfrac{d}{dx}$ satisfying the Heisenberg relation. To get a clearer mathematical presentation one can replace the operator $x$ with $C^\infty_c (\R)$, the space of smoothly supported compact function, and the operator $ i \frac{d}{dx}$ with translations in $\R$, i.e. by the operators
$$ U_{a}\xi (x)=\xi (x- a ) , \quad a\in \R, \xi \in L^2(\R) .$$
The advantage of this latter picture are multiple, one being that the representation theory is clearer.

With the $\mathbf{QHD}(M)$ algebra we adopt precisely this setup -- i.e. functions on a configuration space together with translation operators -- to formulate a framework for a non-perturbative quantum field theory for gauge fields. 
In gauge field theories the underlying configuration space is the space $\ca$ of connections with values in a chosen structure group. 
 We therefore need an algebra of 'functions' on $\ca$. The $\mathbf{QHD}(M)$ algebra includes an algebra of operator valued functions on $\ca$, which is generated by holonomies along flows of vector fields on the manifold. The translation part of the $\mathbf{QHD}(M)$ algebra is obtained by noting that two connections in $\ca$ differs by a 1-form with values in the Lie algebra of the structure group. Therefore for each Lie algebra valued one-form in the manifold we have a  translation operator on $\ca$.

Once we have the $\mathbf{QHD}(M)$ algebra a critical question -- in particular for an application to physics -- is if there exists reasonable representations of it. Hilbert space representations form the foundation of quantum theories and hence the question whether they exist is of paramount importance.
In \cite{MR3562665} we gave an informal argument that no reasonably local representations of the $\mathbf{QHD}(M)$ algebra exists. Local in this context means the following: If $\omega$ is a 1-form and if we partition the underlying manifold into two sets, then the transition probability of $\omega$ on the entire manifold is the product of the transition probability of $\omega$ on each of the subsets. 

The problem with this sort of locality is that  there is no correlation between neighbouring points, which in turn means that fast oscillating connections have the same probability as slow oscillating connections. This renders the expectation value of a holonomy-diffeomorphism on every path, apart from the trivial one, zero.   

Therefore, in this paper we give up the requirement of locality in order to construct representations. We do this by demanding that fast oscillating connections, or actually fast oscillating transitions between connections, have smaller probability than slow oscillating transitions. 

Concretely, for the rate of oscillation we choose the eigenvalues of the Hodge-Laplace operator. We then expand the 1-forms in the eigenvectors of the Hodge-Laplace operator and construct a Gaussian measure on the coefficients of this expansion weighted with the corresponding eigenvalue of the Hodge-Laplace operator. This construction bears some resemblance to the approach by Jaffe et al. in \cites{Glimm, Jaffe}.

The result of this  construction is that the transition probability of a 1-form $\omega$ depends on the Sobolev norm of $\omega$. The construction is therefore not local, since for a given  partition of the underlying manifold into two disjoint sets, splitting $\omega$ into a sum $\omega_1 +\omega_2$, where $\omega_1$ and $\omega_2$ have disjoint support in the sets, would typically render $\omega_1$ and $\omega_2$ discontinuous. Their Sobolev norm would therefore be infinite, i.e. the transition probabilities would be zero. 

One important feature of the representation is that it is strongly continuous. In particular this means that we can define infinitesimal operators $\hat{\nabla}_X$ and $E_\omega$ -- a quantised covariant derivative and an infinitesimal translation operator -- which reproduces the structure of the canonical commutation relations of a Yang-Mills theory. Since the infinitesimal translation operator $E_\omega$ cannot be arbitrarily localised, the canonical commutation relations are only local up to a scale, which we tentatively interpret as the Planck scale.

Once we have infinitesimal operators we devise a general method of constructing physically interesting operators such as the Hamilton operator of a Yang-Mills theory.

One major drawback of the representations, which we find, is that they break the gauge symmetry. This feature, which seems to jeopardise the soundness of our entire enterprise, is a direct consequence of the Hodge-Laplace operator and the Sobolev norm, since the ultra-violet regulation, which it brings about, is not gauge covariant. There are two reasons, why we nevertheless believe that our results are worth considering:
\begin{enumerate}
\item
separable and strongly continuous Hilbert space representations of algebras, that encode the canonical commutation relations of field theories in arbitrary dimensions are exceedingly rare. We are in fact not aware of any such result in 3 + 1 dimensions.
\item
in \cite{A4} we have presented a framework in which the ultra-violet regularisation, which the Hodge-Laplace operator effectuates, is indeed covariant. In that case, however, we do not yet have a complete proof that the corresponding Hilbert space representation exist. The present paper serves, therefore, as a stepping-stone towards a full proof of a separable and strongly continuous Hilbert space representation of the $\mathbf{QHD}(M)$ algebra that is gauge covariant.
\end{enumerate}

One important advantage of the approach to non-perturbative quantum field theory, that we present, is that it works on curved backgrounds (see for instance \cite{Jaffe} for another, similar, approach). Furthermore, in \cites{AAA2,A4} we show that the Hamilton operator of a Yang-Mills theory coupled to a fermionic sector emerges in a local and flat limit from an infinite-dimensional Bott-Dirac operator, which interacts with the $\mathbf{QHD}(M)$ algebra. This result connects our approach to the usual Fock-space quantisation known from perturbative quantum field theory on flat backgrounds. The plane wave expansion, which is usually the starting point in perturbative quantum field theory, is in \cites{AAA2,A4} shown to emerge from an expansion of field operators in terms eigenfunctions of the Hodge-Laplace operator together with a corresponding operator expansion of the translation operators on the configuration space of connections.

Quantum field theory is generally speaking founded on the basic principles of locality and Lorentz invariance. In the axiomatic approaches, for instance, these principles are encoded in the Osterwalder Schrader axioms \cite{Osterwalder} for the Euclidean theory and in the G\.{a}rding-Wightman \cite{Wightman} or the Haag-Kastler \cites{Haag, Haag2} axioms for the Lorentzian theory. The framework that we propose will, however, be inherently non-local due to the built-in ultra-violet regularisation: in this paper in a non-covariant manner and in \cite{A4} in a covariant and thus physically more realistic manner. This raises the question whether such a framework will preserve Lorentz invariance and causality. One possibility is that there will be a Planck scale violation of the Lorentz symmetry in the form of a modification with a scale transformation, something that could very well be within experimental bounds \cite{Jacobson}. 

Thus, the approach that we proprose differs from other rigorous approaches to quantum field theory such as algebraic quantum field theory \cites{Bogolyubov:1990kw,Brunetti:2015vmh} (see also \cite{Buchholz:2019rem} for recent results) and axiomatic quantum field theory  \cites{Osterwalder,Wightman,Haag, Haag2} in two major ways: firstly, it does not presume locality and secondly it permits a possible Planck-scale correction to the Lorentz symmetry. Note here that basic arguments combining quantum mechanics and general relativity suggest that an ultra-violet screening at the Planck scale must exist \cite{Doplicher}; what we suggest is that such a screening arises from representation theory of basic operator algebras and not, as is generally expected, from a theory of quantum gravity.

The overall aim \cite{A1,A2} of this research project is to apply the machinery of non-commutative geometry \cite{Connesbook} to functional spaces and thereby attempt to establish a connection to the noncommutative formulation of the standard model \cites{Connes1, Connes2}. The infinite-dimensional Bott-Dirac operator constructed in \cite{AAA2,A4} combined with the $\mathbf{QHD}(M)$ or $\mathbf{HD}(M)$ algebras constitute precisely such a non-commutative geometrical structure on a configuration space of connections. The idea is to interpret quantum field theory in terms of such geometrical structures and therewith obtain a guiding principle to construct non-perturbative theories. The present paper is a part of the foundation of this research project. 

Finally let us mention that the notion of a metric structure on a configuration space of connections is not new  but was discussed already by Feynman \cite{Feynman} and Singer \cite{Singer} (see also \cite{Orland} and references therein). The construction which we propose in \cites{AAA2,A4} is, however, different from what has previously been discussed, both with respect to techniques employed and with respect to its conceptual scope.

\subsection{Outline of the paper}

We first introduce the $\mathbf{QHD}(M)$ algebra in section \ref{QHD}. Since the holonomy-diffeomorphisms are operator valued functions over a configuration space $\ca$ of connections we construct in section \ref{L2} a Hilbert space $L^2(\ca)$ as an inductive limit over finite dimensional spaces. In subsection \ref{uomegal2} we then represent the translation operators hereon and include the holonomies in subsection \ref{holl2}. In subsection \ref{staerktobak} we show that these representations are strongly continuous and in section \ref{full} that everything adds up to a representation of the $\mathbf{QHD}(M)$ algebra. Finally we show in section \ref{hodge} that the Hodge-Laplace operator gives a concrete realisation of the representation, which to this point has depended on the existence of an orthonormal basis of one-forms satisfying a number of requirements. We end the paper in section \ref{physical} with an outline of how physical operators are constructed and with a note on physical interpretation in section \ref{note}.

\section{The quantum holonomy-diffeomorphism algebra}

\label{QHD}

We start with the holonomy-diffeomorphism algebra $\mathbf{H D} (M)  $, which was first introduced in \cite{MR3046487,MR3428354}, and the quantum holonomy-diffeomorphism algebra $\mathbf{QH D} (M) $  as well as its infinitesimal version $\mathbf{dQH D} (M)  $, which  were introduced in  \cite{MR3482818,MR3562665}.

\subsection{The holonomy-diffeomorphism algebra}
\label{beent}

Let $M$ be a compact smooth manifold of dimension $d$, let $G$ be a compact Lie group, and let $\upsilon :G\to M_j (\mathbb{C})$ be a unitary faithful representation. Consider the vector bundle $S=M\times \C^j$ over $M$ as well as the space of $G$ connections acting on the bundle. Given a metric $g$ on $M$ we get the Hilbert space $L^2(M,S,dg)$, where we equip $S$ with the standard inner product. Given a diffeomorphism $\phi:M\to M$ we get a unitary operator $\phi^*$ on  $L^2(M,S,dg)$ via
$$( \phi^* (\xi ))(\phi (x) )= (\Delta \phi )(M)  \xi (x) , $$
where  $\Delta \phi (x)$ is the volume of the volume element in $\phi (x)$ induced by a unit volume element in $ x$ under $\phi $.      

Let $X$ be a vector field on $M$ and let $\nabla$ be a $G$-connection acting on $S$.  Denote by $t\to \exp_t(X)$ the corresponding flow. Given $x\in M$ let $\gamma$ be the curve  
$$\gamma (t)=\exp_{t} (X) (x) $$
running from $x$ to $\exp_1 (X)(x)$. We define the operator 
$$e^X_\nabla :L^2 (M , S, dg) \to L^2 (M ,  S , dg)$$
in the following way:
we consider an element $\xi \in L^2 (M ,  S, dg)$ as a $\C^j$-valued function, and define 
\begin{equation}
  (e^X_\nabla \xi )(\exp_1(X) (x))=  ((\Delta \exp_1) (x))  \hbox{Hol}(\gamma, \nabla) \xi (x)   ,
  \label{chopin1}
 \end{equation}
where $\hbox{Hol}(\gamma, \nabla)$ denotes the holonomy of $\nabla$ along $\gamma$. Note that $e^X$ is a unitary operator. 
Let $\ca$ be the space of $G$-connections acting on $S$. We have an operator valued function on $\ca$ defined via 
\begin{equation}
\ca \ni \nabla \to e^X_\nabla  . 
\nn
\end{equation}
We denote this function $e^X$. For a function $f\in C^\infty_c (M)$ we get another operator valued function $fe^X$ on $\ca$. We call this operator a holonomy-diffeomorphisms. 
Denote by $\cf (\ca , \mathbb{B} (L^2(M, S,dg) ))$ the bounded operator valued functions over $\ca$. This forms a $C^*$-algebra with the norm
$$\| \Psi \| =  \sup_{\nabla \in \ca} \{\|  \Psi (\nabla )\| \}, \quad \Psi \in  \cf (\ca , \mathbb{B} (L^2(M, S,dg )) ). $$

\begin{definition}
Let 
$$C =   \hbox{span} \{ fe^X |f\in C^\infty_c(M), \ X \hbox{ vector field on  } M\}  . $$
The holonomy-diffeomorphism algebra $\mathbf{H D} (M,S,\ca)   $ is defined to be the $C^*$-subalgebra of  $\cf (\ca , \cb (L^2(M,S,dg )) )$ generated by $C$.
We will often denote $\mathbf{H D} (M,S,\ca)   $ by  $\mathbf{H D}  (M)$ when it is clear which $S$ and $\ca$ is meant.
\end{definition}

It was shown in \cite{MR3428354} that  $\mathbf{H D} (M,S,\ca)   $ is independent of the metric $g$. \\

\subsection{The quantum holonomy-diffeomorphism algebra}

Let $\mathfrak{g}$ be the Lie-algebra of $G$.   
A section $\omega \in \Omega^1(M,\mathfrak{g})$ induces a transformation of $\ca$, and therefore an operator $U_\omega $ on $\mathcal{F}(\ca,  \mathbb{B} (L^2 (M ,  S,g)))$ via   
$$U_\omega (\xi )(\nabla) = \xi (\nabla - \omega) ,$$ 
which satisfy the relation 
\begin{equation} \label{konj}
(U_{\omega} fe^X U_{ \omega}^{-1}) (\nabla) = fe^X (\nabla + \omega )  .
\end{equation}
%
%
Infinitesimal translations on $\ca$ are formally given by 
\begin{equation}
E_\omega  =\frac{d}{dt}U_{  t  \omega}\Big|_{t=0} \;,
\label{soevnloes}
\end{equation}
where we note that 
$$
E_{\omega_1+\omega_2}=E_{\omega_1}+E_{\omega_2\;,}
$$
which follows since the map $\Omega^1 (M,\mathfrak{g})\ni \omega \to U_{ \omega}$ is a group homomorphism, i.e. $U_{(\omega_1+\omega_2 )}=U_{\omega_1}U_{ \omega_2}$.

\begin{definition}
We define the $\mathbf{QHD}(M)$ as the algebra generated by elements in $\mathbf{HD}(M)$ and by translations $U_{\omega}$. We define the infinitesimal quantum holonomy-diffeomorphism algebra $\mathbf{dQHD}(M)$ as the algebra generated by elements in $\mathbf{HD}(M)$ and by infinitesimal translations $E_{\omega}$.
\end{definition}

A priory $\mathbf{QHD}(M)$ is not a $*$-algebra. We will however require of a representation, that it makes the $U_\omega$'s unitary. This in turn makes $\mathbf{QHD}(M)$ a $*$-algebra.

Due to the construction of $\mathbf{QHD}(M)$ as an algebra of operator valued functions over $\ca$ together with translations on $\ca$, in order to construct a representation of  $\mathbf{QHD}(M)$, one needs to construct a suitable $L^2(\ca)$, such that the operators $U_\omega$ act as well defined translations on $L^2(\ca)$, and such that the expectation value of the holonomies on a state in $L^2(\ca)$ is well defined.



\section{Construction of the Hilbert space $L^2(\ca )$}

\label{L2}

The construction of the Hilbert space will depend on a choice of a connection $\nabla_0  \in \ca$, as well as some data given below.

Since we have a representation $\upsilon$ of $\mathfrak{g}$, we get a scalar product on $\mathfrak{g}$ via $tr(\upsilon ( g_1^*)\upsilon (g_2))$, $g_1,g_2\in \mathfrak{g}$, where $tr$ denotes the matrix trace.
We denote the fiberwise scalar product   on $T^*M \times  \mathfrak{g}$  induced by $g$ and the scalar product on $\mathfrak{g}$ by   $( \cdot , \cdot )$. 
Furthermore we choose a scalar product $\langle \cdot , \cdot \rangle_{\mathrm{s}}$ on $\Omega^1 (M, \mathfrak{g})$ and a system $\{ e_i \}_{i \in \mathbb{N}}$ of vectors in $\Omega^1 (M, \mathfrak{g})$ with the properties:
\begin{itemize}
\item[(i)]  That $\{ e_i \}_{i \in \mathbb{N}}$ is an orthonormal basis of the completion $\overline{\Omega^1 (M, \mathfrak{g})}$ with respect to $\langle \cdot , \cdot \rangle_{\mathrm{s}}$.
\item[(ii)] That
\begin{equation}
\sum_i^\infty \| e_i\|_\infty^2   <\infty ,
\end{equation}
with $\| e_i\|_\infty =\sup_{m \in M} (e_i(m),e_i(m))   $.  
\label{stein}
\end{itemize}
Note that since $M$ is compact this notion is independent of the choice of $g$.  Also note that $\Omega^1 (M, \mathfrak{g})$ is a real and not a complex vector space.

We put $\ca_n=\nabla_0 + \hbox{span} \{e_1, \ldots , e_n \} $ and identify $\ca_n$ with $\mathbb{R}^n$ via 
$$\Phi (x_1, \ldots , x_n ) = \nabla_0 +  x_1e_1+ \ldots + x_ne_n .$$
 We define $L^2(\ca_n)$ as $L^2 (\mathbb{R}^n)$ under this identification, i.e. for $\xi ,\eta \in L^2 (\ca_n)$ we have 
 $$\langle \xi , \eta  \rangle_{\ca_n} =\int_{-\infty}^\infty \cdots \int_{-\infty}^\infty \overline{\xi (\nabla_0 + x_1e_1+\ldots + x_ne_n)} {\eta (\nabla_0 + x_1e_1+\ldots + x_ne_n)}dx_1 \cdots dx_n .$$
There is an embedding of Hilbert spaces  $\iota_{n,n+1} :L^2 (\ca_n) \hookrightarrow L^2 (\ca_{n+1}) $ defined as
$$\iota_{n,n+1} (\xi)(\nabla_0 + x_1e_1+\ldots + x_ne_n+x_{n+1}e_{n+1}) =\xi ( \nabla_0 + x_1e_1+\ldots + x_ne_n) \frac{1}{\sqrt[4]{\pi}} e^{-\frac{x_{n+1}^2}{2}} .$$
We also denote $\iota_{n,n+1}(L^2 (\ca_n))$ with $L^2 (\ca_n)$.

\begin{definition}
We define
$$L^2 (\ca) =\lim_{n\to \infty} L^2 (\ca_n) ,$$
as the inductive limit Hilbert space  of the sequence 
$$L^2 (\ca_1) \stackrel{\iota_{1,2}} \longrightarrow L^2 (\ca_2) \stackrel{\iota_{2,3}} \longrightarrow L^2 (\ca_2)\stackrel{\iota_{3,4}} \longrightarrow \ldots  $$
We denote the scalar product on $L^2 (\ca)$ by $\langle \cdot , \cdot \rangle_{\ca}$.

We get embeddings $\iota_n :L^2(\ca_n)\hookrightarrow L^2(\ca)$ of Hilbert spaces. We will also denote $\iota_n(L^2 (\ca_n))$ with $L^2 (\ca_n)$. Furthermore we define 
$$L^2(\ca)_{\mathrm{ alg}}=\bigcup_{n\in \mathbb{N}} L^2(\ca_n).$$

The state in $L^2(\ca )$ of the form 
$$\Phi (\nabla_0 +x_1e_1+\ldots + x_ne_n+\ldots)=\pi^{-\frac{1}{4}}e^{-\frac{x_1^2}{2}}\cdots \pi^{-\frac{1}{4}}e^{-\frac{x_n^2}{2}} \cdots $$
will be called the ground state.

\end{definition}

\subsection{The operators $U_\omega$ on $L^2(\ca )$}
\label{uomegal2}

The operators $U_\omega$ will act on $L^2 (\ca )$ simply by translation, i.e. formally by $(U_\omega \xi) (\nabla)=\xi (\nabla-\omega) $. To make this definition precise we expand 
$$\omega =\sum_{i=1}^\infty  a_i e_i $$
with $a_i= \langle \omega , e_i  \rangle_{\mathrm{s}}$. 
Put $\omega_k=\sum_{i=1}^k  a_i e_i$. On $L^2 (\ca_n)$, $n\geq k$, $U_{\omega_k}$ acts as
\begin{eqnarray*} 
\lefteqn{(U_{\omega_k} (\xi))(\nabla_0+x_1e_1+\ldots + x_ne_n ) }\\
&  = \xi (\nabla_0 +(x_1-a_1)e_1+(x_2-a_2)e_2+\ldots + (x_k-a_k)e_k +x_{k+1}e_{k+1} +\ldots x_ne_n ).
\end{eqnarray*}
Note that this action is compatible with $\iota_{n,n+1}$. It therefore follows that $U_{\omega_k}$ defines an operator on $L^2(\ca)_{\mathrm{alg}}$. Furthermore, since $U_{\omega_k}$ acts as a translation, it is a unitary operator with $U_{\omega_k}^*=U_{-\omega_k}$ and $U_{\omega_k}U_{\nu_k}=U_{\omega_k+\nu_k}$. Hence $U_{\omega_k}$ also extends uniquely to a unitary operator on $L^2(\ca)$. 

Now let $\xi , \eta \in L^2(\ca_n)$ but this time with $k\geq n$. A small computation gives 
\begin{eqnarray*}
\langle \eta , U_{\omega_k} \xi \rangle_{\ca}  &=& \int_{-\infty}^\infty \cdots \int_{-\infty}^\infty   
\overline{\eta (\nabla_0 +x_1 e_1+\ldots x_n e_n )}  \\
&&\times  \xi (\nabla_0+(x_1-a_1)e_1+(x_2-a_2)e_2+(x_n-a_n)e_n)  \\
&& \times  \frac{1}{\sqrt{\pi}^{k-n}} e^{-\frac{1}{2}\left( x_{n+1}^2 +(x_{n+1}-a_{n+1})^2 + \ldots   x_{k}^2 +(x_{k}-a_{k})^2 \right)}dx_1\cdots dx_{k} \\
&=& \langle \eta , U_{\omega_n} \xi \rangle_{\ca_n} e^{-\frac{1}{4}( a_{n+1}^2+\ldots +a_{k}^2)} .
\end{eqnarray*}
It follows 
$$ \lim_{k\to \infty}  \langle \eta , U_{\omega_k}\xi \rangle_{\ca}   =\langle \eta ,U_{\omega_n} \xi \rangle_{\ca_n} e^{-\frac14 \sum_{n+1}^\infty a_i^2 }.$$
If we thus define 
$$ s(\eta ,\xi )=\lim_{k\to \infty}  \langle \eta , U_{\omega_k}\xi \rangle  ,$$
we get that $s$ is a bounded sesquilinear form on $L^2(\ca)_{\mathrm{alg}}$. Hence $s$ has a unique extension to a bounded sesquilinear  form on $L^2(\ca)$. As such, $s$ uniquely defines an operator $U_\omega :L^2(\ca) \to L^2 (\ca)$. Due to the properties of $U_{\omega_k}$ described above, we have $U^*_\omega=U_{-\omega}$ and $U_{\omega}U_{\nu}=U_{\omega + \nu}$. In particular $U_\omega$ is unitary.

\begin{prop}

\label{staerkomega}

The map 
$$ \Omega^1(M,\mathfrak{g}) \ni \omega \mapsto U_\omega \in \mathbb{B} (L^2 (\ca ))$$
defines a strongly continuous additive map, if we consider $ \Omega^1(M,\mathfrak{g})$ equipped with the topology arising from $\langle \cdot , \cdot \rangle_{\mathrm{s}}$.

\end{prop}

\textit{Proof.} Since $U_\omega$ is unitary, it suffices to prove strong continuity on vectors in $L^2(\ca )_{\mathrm{alg}}$. And since the map is additive, it suffices to prove
$$ \lim_{\omega \to 0}\langle \xi ,U_{\omega} \xi \rangle_{\ca} =  \langle \xi , \xi \rangle_{\ca}  $$
for $\xi \in L^2(\ca )_{\mathrm{alg}}$. For  $\xi \in L^2(\ca_n)$ we have, however, that 
$$ \langle \xi , U_\omega \xi \rangle_{\ca} =\langle \xi ,U_{\omega_n} \xi \rangle_{\ca} e^{-\frac14 \sum_{n+1}^\infty a_i^2 }.$$ 
When $\omega \to 0$ we have $\omega_n \to 0$ and $\lim_{\omega \to 0} \sum_{n+1}^\infty a_i^2=0  $. It follows that $\langle \xi ,U_{\omega_n} \xi \rangle_{\ca } \to \langle \xi , \xi \rangle_{\ca }$, and
that  $e^{-\frac14 \sum_{n+1}^\infty a_i^2 } \to 1$.  {\tiny\qed}
\\
 

\subsection{The holonomies on $L^2(\ca )$}

\label{holl2}

Let $p$ be a path on $M$ parameterised by $\gamma:[a,b]\to M $, and assume that this parametrisation is by arc length with respect to the metric $g$. We want to define the expectation value  
of the holonomy with respect to a vector $\xi \in L^2 (\ca)$ via
\begin{eqnarray*}
\rho_\xi ( p ) 
&=& \lim_{k\to \infty} \int_{-\infty}^\infty \cdots \int_{-\infty}^\infty \hbox{Hol} (p, \nabla_0 +x_1e_1+\ldots +x_k e_k) \\
&&\qquad\qquad \times  |\xi (\nabla_0 +x_1 e_1+\ldots +x_ke_k)|^2dx_1 \cdots dx_k   .
\end{eqnarray*}
The question is of course, if the expression converges. This is where the condition $\sum_i^\infty \| e_i\|_\infty^2   <\infty $ becomes crucial. This condition basically ensures that the corrections to the expectation value becomes small as $k$ becomes big. 

To prove this, we start by proving it for ground state.  

Note that we have the following asymptotic expansion of the Gaussian integral:
 
$$ \int_c^\infty e^{-x^2}dx\sim \frac{e^{-c^2}}{c}.$$
We choose $c_k^2=2\log k$. We thereby have $\frac{e^{-c_k^2}}{c_k}\leq \frac{1}{k^2}  $. We write 
\begin{eqnarray}
\lefteqn{\frac{1}{\sqrt{\pi}} \int_{-\infty}^\infty \hbox{Hol}(\gamma ,\nabla+x_ke_k)e^{-x_k^2} dx_k }  \nn \\
&=&\frac{1}{\sqrt{\pi}} \int_{-c_k}^{c_k} \hbox{Hol}(\gamma ,\nabla+x_ke_k)e^{-x_k^2} dx_k \nn \\
&& + \frac{1}{\sqrt{\pi}} \int_{-\infty}^{-c_k} \hbox{Hol}(\gamma ,\nabla+x_ke_k)e^{-x_k^2} dx_k  \nn \\
&& + \frac{1}{\sqrt{\pi}} \int_{c_k}^{\infty} \hbox{Hol}(\gamma ,\nabla+x_ke_k)e^{-x_k^2} dx_k . \label{acc}
\end{eqnarray} 
The matrix norm of the two last terms is by construction smaller than $\frac{2}{k^2}$. This estimate is independent of $\nabla$. 
We expand the holonomy
\begin{eqnarray}
\hbox{Hol}(p , \nabla+  \omega ) =\hspace{-2cm}&&
\nn\\&&
\hbox{Hol}(p, \nabla) +  \int_a^b \hbox{Hol}(p_1,\nabla) \omega (\gamma'(t)) \hbox{Hol}(p_2,\nabla) dt
\nn   \\&&
+\frac{1}{2}   \int_a^b  \int_a^b  \hbox{Hol}(p_1,\nabla)  \omega (\gamma'(t_1))  \hbox{Hol}(p_2,\nabla)  \omega (\gamma' (t_2)) \hbox{Hol}(p_3, \nabla) dt_1dt_2  
\nn   \\
&&+\ldots \label{ent} 
\end{eqnarray}
where in the second line $p=p_1\circ p_2$ is a partition of the path $p$ so that $p_1:[a,t] \rightarrow M$ and $p_2:[t,b]\rightarrow M$. Likewise in the third line of (\ref{ent}), where $p=p_1\circ p_2 \circ p_3$ is partitioned according to $p_1:[a,t_1] \rightarrow M$, $p_2:[t_1,t_2] \rightarrow M$ and $p_3:[t_2,b] \rightarrow M$.

We have the estimates
\begin{eqnarray}  
\left\| \int_a^b  \hbox{Hol}(p_1,\nabla) \omega (\gamma'(t)) Hol(p_2,\nabla) dt \right\| & \leq & \|\omega \|_\infty  | p| \nonumber \\
 \bigg\| \frac{1}{2}   \int_a^b  \int_a^b  \hbox{Hol}(p_1,\nabla)  \omega (\gamma' (t_1))  \hbox{Hol}(b_2,\nabla) 
 \qquad&&\nn\\ 
 \times \omega (\gamma' (t_2)) \hbox{Hol}(b_3, \nabla)  dt_1dt_2   \bigg\| & \leq &  \frac{1}{2}\|\omega \|_\infty^2  |p |^2  \nonumber \\
\label{est4}& \vdots & 
\end{eqnarray}
where $|p|$ the length of $p$, and where $\| \cdot \|$ denotes the matrix norm. We can therefore estimate
\begin{eqnarray*}
\lefteqn{ \|\hbox{Hol}(p , \nabla+  \omega )-\big( \hbox{Hol}(p, \nabla) +  \int_a^b  \hbox{Hol}(p_1,\nabla) \omega (\gamma' (t)) \hbox{Hol}(p_2,\nabla)dt
\nn} \\&&
+\frac{1}{2}   \int_a^b  \int_a^b   \hbox{Hol}(p_1,\nabla)  \omega (\gamma'(t_1))  \hbox{Hol}(p_2,\nabla) 
 \omega (\gamma' (t_2)) \hbox{Hol}(p_3, \nabla)   \big)  dt_1dt_2   \|\\
& \leq & e^{\|\omega \|_\infty  | p |}-\left( 1+\|\omega \|_\infty  |p |+\frac12 \left( \|\omega \|_\infty  |p |\right)^2 \right)
\end{eqnarray*}
The Taylor formula gives the following estimate
$$\left| e^{\|\omega \|_\infty  |p |}-\left( 1+\|\omega \|_\infty  |p |+\frac12 \left( \|\omega \|_\infty  |p |\right)^2 \right)\right| \leq \frac16 (\|\omega \|_\infty  |p |)^3e^{\|\omega \|_\infty  |p |}$$
For $\|\omega \|_\infty  |p | \leq 1$ we therefore have 
\begin{eqnarray}
\lefteqn{ \|  \hbox{Hol}(p , \nabla+  \omega )-\big(  \hbox{Hol}(p, \nabla) +  \int_a^b   \hbox{Hol}(\gamma_1,\nabla) \omega (\gamma' (t)) Hol(\gamma_2,\nabla) dt
\nn} \\&&
+\frac{1}{2}   \int_a^b  \int_\gamma    \hbox{Hol}(\gamma_1,\nabla)  \omega (\gamma' (t_1))   \hbox{Hol}(\gamma_2,\nabla)  \omega ( \gamma' (t_2))  \hbox{Hol}(\gamma_3, \nabla)   \big) dt_1dt_2    \|  \nonumber \\
& \leq & \frac{1}{2}(\|\omega \|_\infty  | p |)^3. \label{est}
\end{eqnarray}
We remind the reader of the following Gaussian integrals:
$$\frac{1}{\sqrt{\pi}}\int_{-\infty}^\infty (ax)^2e^{-x^2} dx=\frac{a^2}{2} $$
$$\frac{1}{\sqrt{\pi}}\int_{-\infty}^\infty |ax|^3e^{-x^2} dx=a^3 . $$
We choose $n$  big enough with 
\begin{equation}
\|e_n\|_\infty |p | c_n\leq \frac{1}{n^{\frac{5}{12} }} \label{est1}
\end{equation}
 We can do this, since $c_n^2=2\log (n)$ and  $\sum_i^\infty \|e_n\|_\infty^2 <\infty$. Note that the power $\frac{5}{12}$ is not fundamental but simply chosen for the following estimate to work.

For $k\geq n$ we rewrite as follows:
\begin{eqnarray*}
&& \pi^{-\frac{k}{2}}\int_{-\infty}^\infty \cdots\int_{-\infty}^\infty \int_{-\infty}^{\infty} \hbox{Hol} (p , \nabla_0 +x_1e_1 +\ldots x_ke_k) e^{-(x_1^2+\ldots x_k^2)}dx_1 \cdots dx_k  \\
&=& \pi^{-\frac{k}{2}}\int_{-\infty}^\infty \cdots\int_{-\infty}^\infty \int_{-c_k}^{c_k} \hbox{Hol} (p , \nabla_0 +x_1e_1 +\ldots x_ke_k) e^{-(x_1^2+\ldots x_k^2)}dx_1 \cdots dx_k  \\
&& +\pi^{-\frac{k}{2}}\int_{-\infty}^\infty \cdots\int_{-\infty}^\infty \int_{-\infty}^{-c_k} \hbox{Hol} (p , \nabla_0 +x_1e_1 +\ldots x_ke_k) e^{-(x_1^2+\ldots x_k^2)}dx_1 \cdots dx_k  \\
&& + \pi^{-\frac{k}{2}}\int_{-\infty}^\infty \cdots\int_{-\infty}^\infty \int_{c_k}^{\infty } \hbox{Hol} (p , \nabla_0 +x_1e_1 +\ldots x_ke_k) e^{-(x_1^2+\ldots x_k^2)}dx_1 \cdots dx_k .
\end{eqnarray*}
We can estimate the last two integrals with $\frac{2}{k^2}$. 
We introduce the following notation:
$$x=(x_1, \ldots , x_{k-1}), \quad x^2=x_1^2+\ldots +x_{k-1}^2 ,$$
$$dx=dx_1 \cdots dx_{k-1} , $$
$$H_p(x)= \hbox{Hol} (p , \nabla_0 +x_1e_1 +\ldots x_{k-1}e_{k-1}) ,$$
and
$$ H_p(x,x_k)= \hbox{Hol} (p , \nabla_0 +x_1e_1 +\ldots x_{k-1}e_{k-1} + x_{k}e_{k}).$$

The first integral can be rewritten:
\begin{eqnarray*}
&& \pi^{-\frac{k}{2}}\int_{-\infty}^\infty \cdots\int_{-\infty}^\infty \int_{-c_k}^{c_k} e^{-(x^2+ x_k^2)} H_p(x,x_k)\  dx\  dx_k  \\
&=& \pi^{-\frac{k}{2}}\int_{-\infty}^\infty \cdots\int_{-\infty}^\infty \int_{-c_k}^{c_k} e^{-(x^2+ x_k^2)}\bigg(  \Big(  H_p(x,x_k) \\
&&   -H_p(x)  -  \int_a^b  H_{p_1}(x)  x_{k}e_k(\gamma'(t)) H_{p_2}(x)\ dt \\
&& -\frac12 \int_a^b\int_a^b H_{p_1}(x)x_ke_k(\gamma' (t_1)) H_{p_2}(x)x_ke_k (\gamma'(t_2))H_{p_3}(x) dt_1dt_2  \Big)\\
&&+  H_p(x)  +  \int_a^b  H_{p_1}(x)  x_{k}e_k(\gamma'(t)) H_{p_2}(x)\ dt \\
&& +\frac12 \int_a^b\int_a^b H_{p_1}(x)x_ke_k(\gamma' (t_1)) H_{p_2}(x)x_ke_k (\gamma'(t_2))H_{p_3}(x) dt_1dt_2 \bigg) dxdx_k
\end{eqnarray*}
The first part of the integrand, i.e. 
\begin{eqnarray*}
&& H_p(x,x_k)   -H_p(x)  -  \int_a^b  H_{p_1}(x)  x_{k}e_k(\gamma'(t)) H_{p_2}(x)\ dt \\
&& -\frac12 \int_a^b\int_a^b H_{p_1}(x)x_ke_k(\gamma' (t_1)) H_{p_2}(x)x_ke_k (\gamma'(t_2))H_{p_3}(x) dt_1dt_2 
\end{eqnarray*}
can with the help of (\ref{est}) ($\nabla =\nabla_0 +x_1e_1 +\ldots x_{k-1}e_{k-1}$) be estimated  with 
\begin{eqnarray*}
\pi^{-\frac{k}{2}}\int_{-\infty}^\infty \cdots\int_{-\infty}^\infty \int_{-c_k}^{c_k}  e^{-(x^2+ x_k^2)} (\|e_k\|_\infty|p |)^3|x_k|^3 dx  dx_k \leq \frac{1}{k^{\frac{5}{4}}}
\end{eqnarray*}
since  $|x_k|\|e_k\|_\infty | p |\leq \|e_k\|_\infty | p | c_k \leq \frac{1}{k^{\frac{5}{12} }} $ according to (\ref{est1}).

In the second integrand we have the term
$$\int_\gamma H_{p_1} (x) x_{k}e_k(\gamma' (t)) H_{p_2}(x) dt .$$
This is odd in $x_k$ and hence vanishes when integrated over $x_k$. The last term  can be estimated with  
\begin{eqnarray*}
\big\| \frac{1}{2}   \int_a^b  \int_a^b   H_{p_1}(x)  x_{k}e_k(\gamma ' (t_1))  H_{p_2}(x)   x_{k}e_k(\gamma ' (t_2)) H_{p_3}(x)\ dt_1 dt_2 \big\| \\  
 \leq  \frac{1}{2} (\|e_k\|_\infty| p |)^2 x_k^2,
 \end{eqnarray*}
 and after multiplying with $e^{-x_k^2}$ and integrating with respect to $x_k$ over $[-c_k,c_k]$ the term can be estimated with  $\frac{1}{2} (\|e_k\|_\infty| p |)^2$. 
All together we have
\begin{eqnarray*}
&& \frac{1}{{\pi}^{\frac{k}{2}}}\int_{-\infty}^\infty \cdots\int_{-\infty}^\infty \hbox{Hol} (\gamma , \nabla +x_1e_1 +\ldots x_ke_k) e^{-(x_1^2+\ldots x_k^2)}dx_1 \cdots dx_k \\
&& =\frac{1}{{\pi}^{\frac{k}{2}}}\int_{-\infty}^\infty \cdots\int_{-\infty}^\infty \hbox{Hol} (\gamma , \nabla +x_1e_1 +\ldots x_{k-1}e_{k-1}) e^{-(x_1^2+\ldots x_{k-1}^2)}dx_1 \cdots dx_{k-1} 
\end{eqnarray*}
with an error smaller than $\frac{2}{k^2}+\frac{1}{2} (\|e_k\|_\infty| p |)^2+\frac{1}{k^{\frac{5}{12} }}$.  The series $\sum_{k=1}^\infty \|e_k\|_\infty^2$ is convergent, and hence the  series 
$$\sum_{k=1}^\infty\left(  \frac{2}{k^2}+\frac{1}{2} (\|e_k\|_\infty| p |)^2+\frac{1}{k^{\frac{5}{4} }} \right)$$ 
is also convergent.  We thus have
\begin{prop} \label{limit}
The limit  
$$\lim_{k\to \infty}\frac{1}{{\pi}^{\frac{k}{2}}} \int_{-\infty}^\infty \cdots\int_{-\infty}^\infty  \hbox{Hol} (\gamma , \nabla +x_1e_1 +\ldots x_ke_k) e^{-(x_1^2+\ldots x_k^2)}dx_1 \cdots dx_k$$
exist.
\end{prop}

We now want to extend this to general  $\xi \in  L^2 (\ca )$. To this end we consider $\mathcal{H}_j=L^2(\mathcal{A})\otimes \mathbb{C}^j$. We think of elements in  $\mathcal{H}_j$ as functions on  $\mathcal{A}$ with values in  $\mathbb{C}^j$. 

Given $p$ we want to define the operator $h_p$ on $\mathcal{H}_j$. The idea is to define 
$$(h_p \xi )(\nabla )=\hbox{Hol} (p , \nabla )\xi (\nabla )  .$$
We have to make sure that this is well defined. For a start assume that $\xi , \eta \in \mathcal{H}_{\mathrm{alg},j}=L^2(\mathcal{A})_{\mathrm{alg}}\otimes \mathbb{C}^j$. It follows from the computation leading to proposition $\ref{limit}$ that $\langle \eta ,h_p \xi \rangle  $ exists, since the integral converges in the first finitely many variables, and the rest of the variables we can control with  
$$\sum_{k=1}^\infty\left(  \frac{2}{k^2}+\frac{1}{2} (\|e_k\|_\infty| p |)^2+\frac{1}{k^{\frac{5}{4} }} \right).$$   
Furthermore  we have $| \langle \eta ,h_p \xi \rangle  | \leq \|\eta\|\|\xi \|$, since it is an integral over expectation values of unitary operators. The expression  
$$(\eta, \xi ) \to \langle \eta ,h_p \xi \rangle $$
is thus a sesquilinear form, and hence uniquely defines an operator $h_p$ on  $\mathcal{H}_j=L^2(\mathcal{A})\otimes \mathbb{C}^j$.

\subsection{Strong continuity}

\label{staerktobak}

For two curves $\gamma_1$ and $\gamma_2$ we define
\begin{equation} \| \gamma_1 -\gamma_2\|_{\mathrm{sob}} = \sup_{t\in [0,1]} ( \|\gamma_1(t)-\gamma_2(t)\| +\|\gamma_1'(t)-\gamma_2'(t) \|).
\label{sobolevcurrve}
\end{equation}
We want to show that 
$$\| \gamma_k -\gamma \|_{\mathrm{sob}} \to 0 \Rightarrow \langle \xi, h_{\gamma_k} \xi \rangle  \to \langle \xi, h_\gamma \xi \rangle. $$ 
This in turn implies strong continuity, i.e. that  
$$\| \gamma_k -\gamma \|_{\mathrm{sob}} \to 0\Rightarrow\| (h_{\gamma_k}-h_\gamma)\xi \| \to 0.$$  

We again start with  $\xi \in \mathcal{H}_{\mathrm{alg},j}=L^2(\mathcal{A})_{\mathrm{alg}}\otimes \mathbb{C}^j$.
Let $\varepsilon >0$  be given. We choose $n$ so big that  
$$\sum_{k=n+1}^\infty\left(  \frac{2}{k^2}+\frac{1}{2} (\|e_k\|_\infty | p |)^2+\frac{1}{k^{\frac{5}{4} }} \right) <\varepsilon ,$$
where $ | p |$ is the arc length of $\gamma$.  We choose $D_n$ with   
$$\int_{(x_1,\ldots , x_n)\notin [-D_n,D_n]\times \cdots  \times [-D_n,D_n]} \|\xi (\nabla_0+x_1e_1+\ldots , x_ne_n) \|_2^2 dx_1 \cdots dx_n < \varepsilon , $$
where $\| \cdot \|_2^2$ denotes the norm squared of a vector in $\mathbb{C}^j$. 
Since $\| \gamma_k -\gamma \|_{\mathrm{sob}} \to 0$ we also have $|\gamma_k| \to |\gamma |$. Furthermore, when $k$ is big enough we have 
\begin{equation}
 \| \hbox{Hol} (\gamma ,\nabla_0 +x_1e_1+\ldots +x_ne_n) - \hbox{Hol} (\gamma_k ,\nabla_0+x_1e_1+\ldots +x_ne_n) \| \leq \varepsilon ,
 \label{klaver}
 \end{equation}
for all $(x_1,\ldots ,x_n)\in [-D_n,D_n]\times \cdots  \times [-D_n,D_n]$. This follows, since we consider a connection  $\nabla_0 +x_1e_1+\ldots +x_ne_n$ along $\gamma$, and the same connection along  $\gamma_k$ as different connections along $\gamma$, and since the map  
$$ [-D_n,D_n]\times \cdots  \times [-D_n,D_n] \ni (x_1,\ldots ,x_n) \to \nabla_0 +x_1e_1+\ldots +x_ne_n$$
is uniformly continuous we can, to a given  $\delta >0$, choose $k$ big enough with  
$$\|  (\nabla_0 +x_1e_1+\ldots +x_ne_n)(\gamma' (t)-\gamma'_k(t) )\| \leq \delta  $$
for all $(x_1,\ldots ,x_n)\in [-D_n,D_n]\times \cdots  \times [-D_n,D_n]$ and $t\in [a,b]$. If we choose  $\delta >0$ small enough we can use the equation (\ref{ent}) together with the estimates (\ref{est4})  to get (\ref{klaver}).

For $k$ big enough we now get:
\begin{eqnarray*}
\lefteqn{|\langle \xi, h_{\gamma_k} \xi \rangle  - \langle \xi, h_\gamma \xi \rangle  |} \\
 &\leq & \big\| \int_{-\infty}^\infty \cdots\int_{-\infty}^\infty \int_{-\infty}^{\infty} \big( \hbox{Hol} (\gamma , \nabla_0 +x_1e_1 +\ldots x_ne_n)  \\
 && -\hbox{Hol}(\gamma_k , \nabla_0 +x_1e_1 +\ldots x_ne_n)  \big)\| \xi (\nabla_0 +x_1e_1+\ldots +x_ne_n) \|^2_2 dx_1 \cdots dx_n\big\| \\
 && + 2 \sum_{k=n+1}^\infty  \left(  \frac{2}{k^2}+\frac{1}{2} (\|e_k\|_\infty| \gamma |)^2+\frac{1}{k^{\frac{5}{4} }} \right) \\
  &\leq & \int_{-D_n}^{D_n} \cdots\int_{-D_n}^{D_n} \int_{-D_n}^{D_n}\big\| \big( \hbox{Hol} (\gamma , \nabla +x_1e_1 +\ldots x_ne_n)  \\
 && -\hbox{Hol}(\gamma_k , \nabla +x_1e_1 +\ldots x_ne_n)  \big) \| \xi (\nabla_0 +x_1e_1+\ldots +x_ne_n) \|^2_2  \big\| dx_1 \cdots dx_n \\
 && +2\varepsilon + 2 \varepsilon  \\
&  \leq & \varepsilon \|\xi\|^2 +4\varepsilon = 5 \varepsilon .
\end{eqnarray*}
We thus have
$\langle \xi, h_{\gamma_k} \xi \rangle  \to \langle \xi, h_\gamma \xi \rangle $.
For general $\xi \in \mathcal{H}_J=L^2(\mathcal{A},j)\otimes \mathbb{C}^j$ we choose to a given   $\varepsilon >0$ a $\xi_n \in \mathcal{H}_{\mathrm{alg},j}=L^2(\mathcal{A})_{\mathrm{alg}}\otimes \mathbb{C}^j$ with $\|\xi_n -\xi \| \leq \varepsilon $ and $\|\xi \|=\| \xi_n\|$. Since $\|h_\gamma \|\leq 1$ for all $\gamma$ it follows that 
$$|\langle \xi |h_\gamma |\xi \rangle - \langle \xi_n |h_\gamma |\xi_n \rangle  | \leq 2\varepsilon \|\xi\| $$
for all $\gamma$. When $k$ is big enough we have 
$|\langle \xi_n |h_\gamma |\xi_n \rangle - \langle \xi_n |h_{\gamma_k} |\xi_n \rangle  | \leq \varepsilon  $, according to what we have just shown. All together we have 
$$ |\langle \xi |h_\gamma |\xi \rangle - \langle \xi |h_{\gamma_k} |\xi \rangle  | \leq \varepsilon+2\|\xi\| \varepsilon  ,$$
i.e. 
$\langle \xi, h_{\gamma_k} \xi \rangle  \to \langle \xi, h_\gamma \xi \rangle $ when 
$\| \gamma_k -\gamma \|_{\mathrm{sob}} \to 0$.
Since $h_\gamma^*=h_{\gamma^{-1}}$ we thus have
\begin{prop} \label{strong1}
The map $\gamma \to h_\gamma \in \mathbb{B} (\mathcal{H}_j )$ ist strongly continuous, i.e.  $$\lim_{k\to \infty} \| \gamma-\gamma_k\| =0\  \Rightarrow  \ \lim_{k\to \infty} \|(h_\gamma -h_{\gamma_k})\xi) \|=0$$ for all $\xi \in \mathcal{H}_j$. 
\end{prop}

\section{The full Hilbert space and the representation of  quantum holonomy-diffeomorphism algebra}

\label{full}

We are now ready to construct the representation of $\mathbf{QHD}(M)$. The Hilbert space we take is
$$\mathcal{H} = L^2(\ca )\otimes L^2(M,S) .$$
The operators $U_{\omega}$ simply act on $\mathcal{H}$ via acting on the  $L^2(\ca )$ component as described in section \ref{uomegal2}. Given a flow $X$ we need to describe how $e^X$ acts on  $\mathcal{H}$. Let $\xi_1,\xi_2 \in L^2(\ca)$ and $\eta_1,\eta_2 \in  L^2(M,S) $. The matrix-valued function 
$$x\to  \langle \xi_1 , h_{\gamma_x} \xi_2 \rangle_\ca ,$$
with $\gamma_x(t)= \exp_t(X)(x)$ is according to proposition \ref{strong1} continuous in $x$. It is thus well defined  to put (compare to (\ref{chopin1})) 
\begin{eqnarray*}
&&\langle \xi_1 \otimes \eta_1 , e^X (\xi_2 \otimes \eta_2 ) \rangle\\
&& = \int_M (\eta_1(\exp_1(x)) ,\Delta \exp_1(x)\langle \xi_1 , h_{\gamma_x} \xi_2 \rangle_{\ca} \eta_2 (x) )_e dg(\exp_1(x)) ,
\end{eqnarray*}
where $(\cdot , \cdot )_e$ denotes the standard inner product on $\mathbb{C}_j$, 

This yields an operator $e^X$ acting on $\mathcal{H}$. With the operators $e^X$ and $U_\omega$ acting on $\mathcal{H}$ we all together have a representation of $\mathbf{QHD}(M)$ on $\mathcal{H}$.

\subsection{The infinitesimals}

According to Proposition \ref{staerkomega} the map $\omega \to U_\omega $ is strongly continuous. It follows from the theorem of Stone, that we have a self-adjoint operator
$$E_\omega = \frac{d}{dt}U_{t\omega} |_{t=0} $$
on $\mathcal{H}$.  Our representation consequently also gives a representation of the $\mathbf{dQHD}(M)$ algebra. 

We now turn to strong continuity of the $e^X$ operators. For this we need a topology on the space of vector fields. We do this by defining 
$$X_n\to X \ \Leftrightarrow \ \| X-X_n\|_\infty \to 0 .$$
Note that if $X_n \to X$ we also have that each integral curve of $X_n$ converges to the corresponding integral curve of $X$ in the norm (\ref{sobolevcurrve}). The convergence is uniform in the choice of start point of the integral curve.     
 
Note that the estimates leading to  strong continuity given in section \ref{staerktobak} only depends on the norm $\| \gamma -\gamma_n\|_{\mathrm{sob}}$. It thus follows:

\begin{prop}
\label{staerkflow}

The map
$$\hbox{Vect} (M)\ni X \to e^X \in \mathbb{B}(\mathcal{H}) $$
is strongly continuous. 
\end{prop}

Putting Proposition \ref{staerkomega} and \ref{staerkflow} together

\begin{thm}

The maps 
$$\hbox{Vect} (M)\ni X \mapsto e^X \in \mathbb{B}(\mathcal{H}) ,$$
and 
$$ \Omega^1(M , \g ) \ni \omega \mapsto  U_\omega \in \mathbb{B}(\mathcal{H}) $$
are strongly continuous, if we consider $\Omega^1(M,\mathfrak{g})$ equipped
with the topology arising from $\langle \cdot ,\cdot \rangle_{\mathrm{s}}$. 
 We consequently have infinitesimal operators
$$\hat{\nabla}_X =\frac{d}{dt}  e^{tX} |_{t=0}  \hbox{ and } E_\omega = \frac{d}{dt}U_{t\omega} |_{t=0} .$$

\end{thm}

Note that $\hat{\nabla}_X$ is, modulo derivatives in the local volume, a quantised covariant derivative.

\section{A special case with the Hodge-Laplace operator}

\label{hodge}
We now address the problem of finding a scalar product $\langle \cdot , \cdot \rangle_{\mathrm{s}}$ on $\Omega^1 (M, \mathfrak{g})$ and a system $\{ e_i \}_{i \in \mathbb{N}}$ of vectors in $\Omega^1 (M, \mathfrak{g})$ with the properties:
\begin{itemize}
\item[(i)]  That $\{ e_i \}_{i \in \mathbb{N}}$ is an orthonormal basis of the completion $\overline{\Omega^1 (M, \mathfrak{g})}$ with respect to $\langle \cdot , \cdot \rangle_{\mathrm{s}}$.
\item[(ii)] That
$$ \sum_i^\infty \| e_i\|_\infty^2   <\infty .$$
\end{itemize}

For the metric $g$ we consider the Hodge-Laplacian 
$$\Delta =  dd^*+d^*d: \Omega^k(M) \to \Omega^k (M),$$
and restrict it to $\Delta : \Omega^1(M) \to \Omega^1 (M)$. Note the following:
\begin{itemize}
\item[1)] The operator is invariant under isometries, 
\item[2)] $\Omega^1(M)$ is a real vector space.  
\end{itemize}

Next we extend the operator to $\Delta: \Omega^1(M,\mathfrak{g} ) \to \Omega^1 (M , \mathfrak{g} )$ by choosing an orthonormal basis for  $\mathfrak{g}$.  
Let $\{ f_i\}_{i\in \mathbb{N}}$ be an orthonormal set  of eigenvectors with eigenvalues $\{ \lambda_i\}_{i\in \mathbb{N}}$ for this operator. Orthonormal here means  with respect to  
$$ \langle u,v \rangle_2 =\int_M  ( u(x), v(x) )    dg(x)  .$$ 
For $u,v\in \Omega^1(M,\mathfrak{g})$ we define 
\begin{equation}
\langle u , v  \rangle_{\mathrm{s}}= \int_M ( (1+\Delta^p)u(x),(1+\Delta^p) v(x) ) dg(x) .
\label{genius}
\end{equation}
We want fix $p$ in order to have condition (ii) for $e_i=\frac{f_i}{\sqrt{\langle f_i , f_i  \rangle_{\mathrm{s}}}}$. 
It follows from the proof\footnote{Private communication with Professor Daniel Grieser.} in  \cite{grieser} that we have
$$\| f_i \|_\infty \sim \lambda_i^\frac{{d-1}}{4}  ,$$
where $d$ is the dimension of $M$, 
and according to  Weyl's asymptotic law the number of eigenvalues smaller than $n$ asymptotically behaves like $n^{\frac{d}{2}}$, or alternatively we have $\lambda_i\sim i^{\frac{2}{d}}$. 
Consequently we have 
$$ \| e_i\|_\infty =\frac{\|f_i\|_\infty }{ 1+\lambda_i^p  } \sim \frac{\lambda_i^{\frac{d-1}{4}}}{1+\lambda_i^p} \sim \lambda_i^{\frac{d-1}{4}-p}\sim i^{\frac{1}{d}\left( \frac{d-1}{2}-2p  \right) },$$
and hence
\begin{prop}

	If $\frac{1}{d}\left( \frac{d-1}{2}-2p  \right) <-\frac12 $, then the conditions (i) and (ii) are fulfilled. 

\end{prop}
For example for $d=3$ the requirement is $p>\frac54$. 

\subsection{Invariance under isometries}

The construction of $L^2(\ca)$ depends on the choice of a basis of eigenvectors of $\Delta$.  Given $\Delta$ this basis is, up to orthogonal transformations of each eigenspace, unique. Since the $n$ dimensional Gaussian integral is invariant under the orthogonal group $O(n)$ it follows that a different choice of basis leads to a unitary transformation of $L^2(\ca )$, which is compatible with the action of the $\mathbf{QHD}(M)$ algebra on $L^2(\ca )$. In this way $L^2(\ca )$ is independent of the choice of basis of eigenvectors of   $\Delta$.

Also the operator $\Delta$ is invariant under isometries of $M$. This makes the construction, up to the choice of $\nabla_0$, invariant under isometries.



\subsection{Support of the measure}

In section \ref{L2} we constructed $L^2(\ca)$. This means that we have a measure on some sort of completion of $\ca$. In this section we will look at this completion and describe it in more detail. 

We first start with the projective limit
$$ \R^\infty =\lim_{\leftarrow } \R^n.$$
We can consider subsets $A\subset \R^n$ as subsets of $\R^\infty$ by mapping 
$$A\mapsto A \times \R \times \R \times \cdots .$$
We denote this map by $\iota_n$.

We have constructed $L^2(\ca)$ by identifying $\ca$ with a subspace of $\R^\infty$ via 
$$
\ca \ni \nabla_0+\sum_{i=1}^\infty x_ie_i \to (x_1,x_2,\ldots, x_n,\ldots )\in \R^\infty   
$$
and considered the measure $\mu$ on  $ \R^\infty $ given for a subset $ A  \subset \R^n$ by\footnote{When we constructed $L^2(\ca)$ we kept the Lebesgue measure in the first finitely many variables in order to write $U_\omega$ in a simple way. }
 $$\mu (\iota_n (A))=\int_A \frac{e^{-(x_1^2+\ldots + x_n^2)}}{\pi^{\frac{n}{2}}} dx_1\cdots dx_n .$$
Let $a>1$. Put $c_n =\sqrt{a\log (n)}$. For $A\subset \R^n$ we define 
$$A_s^a=A\times [-c_{n+1}, c_{n+1}] \times [-c_{n+2}, c_{n+2}] \times \cdots  . $$
We want to compute the measure of this set, or rather we want to show that when $A$ is not a zero set, then $A_s^a$ is also not a zero set. Let $\mu_{G,n}$ be the Gau{\ss} measure in $\R^n$, i.e.
$$\mu_{G,n} (A)=\int_A \frac{e^{-(x_1^2+\ldots + x_n^2)}}{\pi^{\frac{n}{2}}} dx_1\cdots dx_n , \quad A\subset \R^n.$$
 We have  
\begin{eqnarray*}
\lefteqn {\mu (A_s^a)} \\
&=& \lim_{k\to \infty } \mu_{G,n} (A) \frac{1}{\pi^{\frac{k-n}{2}}}\int_{-c_{n+1}}^{c_{n+1}} e^{-x_{n+1}^2} dx_{n+1}\cdots \int_{-c_{n+2}}^{c_{n+2}} e^{-x_{n+2}^2} dx_{n+2} \cdots \int_{-c_{k}}^{c_{k}} e^{-x_{k}^2} dx_{k} .
\end{eqnarray*} 
We can choose $b$ with 
$$\frac{1}{\sqrt{\pi}}\int_{-c_{n+1}}^{c_{n+1}} e^{-x_{n+1}^2} dx_{n+1} \geq 1- b\frac{e^{-c_{n+1}^2}}{c_{n+1}} ,$$
$$\frac{1}{\sqrt{\pi}}\int_{-c_{n+2}}^{c_{n+2}} e^{-x_{n+2}^2} dx_{n+2} \geq 1- b\frac{e^{-c_{n+2}^2}}{c_{n+2}},$$ 
etc.. 
We thus have 
\begin{eqnarray*}
\lefteqn{ \lim_{k\to \infty } \mu_{G,n} (A) \frac{1}{\pi^{\frac{k-n}{2}}}\int_{-c_{n+1}}^{c_{n+1}} e^{-x_{n+1}^2} dx_{n+1} \cdots  \int_{-c_{n+2}}^{c_{n+2}} e^{-x_{n+2}^2} dx_{n+2} \cdots \int_{-c_{k}}^{c_{k}} e^{-x_{k}^2} dx_{k}} \\
 &\geq& \lim_{k\to \infty } \mu_{G,n} (A)\left( 1- b\frac{e^{-c_{n+1}^2}}{c_{n+1}} \right)\left( 1- b\frac{e^{-c_{n+2}^2}}{c_{n+2}} \right) \cdots \left( 1- b\frac{e^{-c_{k}^2}}{c_{k}} \right). 
\end{eqnarray*} 
Taking the logarithm we get: 
$$ \lim_{k\to \infty } \left( 1- b\frac{e^{-c_{n+1}^2}}{c_{n+1}} \right)\left( 1- b\frac{e^{-c_{n+2}^2}}{c_{n+2}} \right) \cdots \left( 1- b\frac{e^{-c_{k}^2}}{c_{k}} \right) \not=0 \ \Leftrightarrow \ \sum_{k=1}^\infty \frac{e^{-c_{k}^2}}{c_{k}}  <\infty .$$
Inserting the value of $c_n$ we get
$$\sum_{k=1}^\infty \frac{k^{-a}}{\sqrt{a\log(k)}} ,  $$
which is a convergent series, and hence we have $\mu (A_s^a) \not= 0$. It follows:
$$ \mu \left(  \bigcup_k(\R^k)_s^a  \right)= 1.$$
We thus have: The support of $\mu$ is contained in the set 
$$ 
S^a =\left\{ (x_1,\ldots , x_n,\ldots )\in \R^\infty \bigg|\ \hbox{there exists }l \hbox{ with } |x_k| \leq \sqrt{a\log(k)} \hbox{ for all }  k\geq l \right\}.
$$

Note, had we chosen $a\leq 1$, the set $S^a$ would be a zero set.  

Let $H^s(M)$ denote the Sobolev space of  weight $s$ of 1-forms with  values in $\mathfrak{g}$. We have the following result
\begin{prop}
Let us assume that we have used the system of eigenvector of the Hodge-Laplace operator to construct $L^2 (\ca)$. The support of the measure induced from the construction of $L^2(\ca)$ is contained in 
$H^s (M) $ for $s<2p-d$, and not contained in $H^s (M) $ for $s\geq 2p-d$.  In fact $H^s (M) $ is a zero set for $s\geq 2p-d$.
\end{prop}

\textit{Proof:}  We basically just need to transport $S^a$ to $\ca$. The support is thus contained in the set of connections of the form 
$$ \nabla_0 +\sum_{k=1}^\infty a_k e_k ,$$
with $\left( \frac{a_k}{\sqrt{a\log(k)}} \right)_{k\in \mathbb{N}}$   bounded.  We have 
$$ e_k=\frac{f_k}{1+\lambda_k^p}.$$
The question is therefore: For which $q$ is
$$ \sum_{k=1}^\infty \sqrt{a \log (k)} \lambda_k^{q-p}$$
convergent. Here the factor $\sqrt{a \log (k)}$ is irrelevant. The asymptotics of $\lambda_k$ is $k^{\frac{2}{d}}$. We thus have $q<p-\frac{d}{2}$. I.e. the support is contained in $H^s (M) $ for $s<2p-d$, and for $s\geq 2p-d$ $H^s(M)$ is a zero set.  {\tiny\qed}

\section{How to construct physically important operators}

\label{physical}




A crucial question is if we with the help of the representation of $\mathbf{QHD}(M)$ can construct operators relevant to quantum field theory, i.e. whether we can construct Hamilton operators for various field theories based on gauge fields\footnote{In \cite{AAA2,A4} we show that the Hamilton operator of a Yang-Mills theory coupled to a fermionic field emerges from an infinite-dimensional Bott-Dirac operator, which interacts with the $\mathbf{QHD}(M)$ algebra. The construction of the Bott-Dirac operator requires, however, additional structure in the form of a CAR algebra. In this section we are concerned with the construction of physically important operators without any additional structure.}. The curvature is in this setting straightforward to quantise, since we have quantised operators $\hat{\nabla}_X$. If, on the other hand, one wants to quantise a classical quantity like 
\begin{equation}
 \mathbf{E}^2=\sum_{ \mu ,  \nu , i}  \int_{M}g_{\m\n} E^\mu_i(x)E^\n_i(x)\   dg(x) ,
 \label{vangelis}
\end{equation}
where $E$ is a vector field that takes values in $\mathfrak{g}$,  then this is not as straightforward to quantise. 
 With the notation used in this article the quantity is
\begin{equation}
 \mathbf{E}^2= \int_{M} \left( E(x) ,E(x) \right) \   dg(x) ,
 \label{vangelis1}
\end{equation}
where we have used the metric to identify vector fields with 1-forms. The problem is that the appearance of the Sobolev norm rules out localising the operators that corresponds to $E_i^\mu(x)$, that is if one wants to localise $E_\omega$ in a point $x$, we need an $\omega$ sharply peaked around $x$. Such an $\omega$ would, however, have a large Sobolev norm, thereby making the expectation value of such an operator small. There exist, however, a quite canonical way to quantise an operator like (\ref{vangelis1}), and in general to approximately localise the $E_\omega$ operators. To this end we consider the heat kernel $K_t(x,y)$, i.e. the kernel of the operator $e^{-t\Delta}$. We consider $K_t(x,y)$ as a function over $M \times M$ with values in $T^*_xM\otimes \mathfrak{g}\otimes T^*_y M\otimes \mathfrak{g}$ at the point $(x,y)$. We denote by $(\cdot , \cdot )_2$ the scalar product in the second variable, i.e. the scalar product in the  $T^*_yM\otimes \mathfrak{g}$ factor. For a vector $v$ in $T^*_yM\otimes \mathfrak{g}$ we consider the function 
$x \to (K_t(x,y),v)_2$. This is an element in  $\Omega^1(M,\mathfrak{g})$, and we hence get the corresponding operator 
$$E_{(K_t(x,y),v)_2}$$
acting on $L^2(\ca)$. Note that since $e^{-t\Delta}  \to 1$ when $t\to 0$ we formally have:
$$E_{(K_t(x,y),v)_2} \to E_{v\delta_y }\hbox{ for }t\to 0 , $$
where $\delta_y$ denotes the delta function in $y$. Thus the limit  $t \to 0$ (ignoring the fact that it does not exists) gives  an operator localised in $y$. We thus have a canonical way of almost localising operators. If we therefore want to quantise (\ref{vangelis1}) we can just take 
$$K^2_t(x_1,x_2)=\int_M (K_t(x_1,y)\otimes K_t (x_2,y))_2 dy   ,$$
where the tensor product means that we tensor $T^*_{x_1}M\otimes \mathfrak{g}$ with $T^*_{x_1}M\otimes \mathfrak{g}$ and define\footnote{There is a linear and continuous (in appropriate topology) map from $\Omega^1 (M,\mathfrak{g})\otimes \Omega^1(M,\mathfrak{g})$  to the operators on $\mathcal{H}$ uniquely defined by  $\omega_1 \otimes \omega_2 \mapsto E_{\omega_1}E_{\omega_2}$, $\omega_1,\omega_2 \in  \Omega^1(M,\mathfrak{g})$. 
We will omit the technical details for now.
}
\begin{equation} \label{localprod}
\hat{\mathbf{E}}^2_t =E_{K^2_t (x_1,x_2)} .
\end{equation}
Note that in terms of the basis $\{ f_k\}$ we have $K_t(x,y)=\sum_{k=1}^\infty e^{-t\lambda_k}f_k(x) \otimes f_k(y)$. In this case the formula reads 
$$\hat{\mathbf{E}}^2_t= \sum_{k,l=1}^\infty e^{-t(\lambda_k+\lambda_l)}\int_M  E_{f_k}E_{f_l} (f_k(y) ,f_l(y)) dy  .$$
On the other hand in our case it is natural to  consider
$$\hat{\mathbf{E}}^2=\sum_{k,l=1}^\infty \int_M E_{e_k}E_{e_l} (e_k(y) ,e_l(y)) dy  . $$
This would correspond to consider $g(-\Delta )$ instead $e^{-t\Delta}  $, with $g(x)=\frac{1}{1+x^p}$. 

 If we let $t\to 0$ in (\ref{localprod}) then the products defined using this type of operators do become local. If we for instance interpret $t$ as the Planck length (see \cite{AAA1} for details), then the products become local up to the Planck length. Also operators constructed in this way are invariant under isometries. 

With this construction we can now built physically interesting operators, for instance Hamilton operators for various gauge theories.
Consider for example the Yang-Mills Hamiltonian, which has the form
\begin{equation}
H_{\mbox{\tiny YM}} = \frac{1}{2}\int d^3x \left( (E^a_\m)^2 + (B^a_\m)^2  \right)
\label{bee}
\end{equation}
where $E$ is the conjugate field to the gauge field $A$, i.e. 
$$
\{ E_\m^a(x), A^b_\n(y)\}_{\mbox{\tiny Poisson}} = \d^{(3)}(x-y) g_{\m\n} \d^{ab}
$$
where $\{\cdot,\cdot\}_{\mbox{\tiny Poisson}} $ is the Poisson bracket. Also,  $B_\m=  \e_\m^{\;\;\n\s} F_{\n\s}$ where $F$ is the field strength tensor of the gauge field $A$. 
There exist therefore a natural candidate for a Yang-Mills Hamilton operator, namely
$$
\hat{H}_{\mbox{\tiny YM}} = \hat{\mathbf{E}}^2 + \hat{\mathbf{F}}^2(\hat{\nabla})
$$
where $\hat{\mathbf{F}}(\hat{\nabla})$ is a curvature operator. This Hamilton operator together with the representation of the $\mathbf{QHD}(M)$ algebra constitute a non-perturbative quantum Yang-Mills theory on $M$. In a similar manner we can also construct the Hamiltonian for general relativity when formulated in terms of Ashtekar connections, see \cite{AAA1} for details.

\section{A note on physical interpretations}
\label{note}

Let us end with a note on the possible physical interpretations of the results obtained. First of all, the representations of the $\mathbf{QHD}(M)$ algebra, which we have identified, involve what amounts to an ultra-violet regularisation in the sense that the scalar product $\langle \cdot , \cdot \rangle_{\mathrm{s}}$ in (\ref{stein}) -- and more specifically the Sobolev norm in (\ref{genius}) -- dampens modes in the ultra-violet. 
Note, however, that this ultra-violet regularisation is of a somewhat different nature compared to what is usually encountered in quantum field theory since it emerges as an integral part of the representation theory of the $\mathbf{QHD}(M)$ algebra and since it does not break any spatial symmetries.

In our opinion this leaves us with two possible interpretations:
\begin{enumerate}
\item
we may interpret the scalar product $\langle \cdot , \cdot \rangle_{\mathrm{s}}$ in (\ref{stein}) and specifically the Laplace operator in (\ref{genius}) in terms of an unphysical cut-off that needs to be removed. This would be in line with ordinary quantum field theory. 
\item
we may interpret the aforementioned cut-off as a physical feature of the theory at hand.
\end{enumerate}
If we choose the first interpretation the question is whether our framework may be useful to provide insights in ordinary quantum field theory and in particular whether it provides a new viable framework for the formulation of  quantum field theories on curved backgrounds. If, on the other hand, we choose the second interpretation then we need to address two additional questions:
\begin{itemize}
  \item[-]
 what happens to the Lorentz symmetry?
 \item[-]
what about the broken gauge symmetry?
 \end{itemize}
Here the second question is the most critical since a broken gauge symmetry jeopardises the framework altogether. In \cite{A4} we have devised a method in which the ultra-violet screening is gauge-covariant. Roughly speaking this is accomplished by using a covariant Hodge-Laplace operator, which promotes the Sobolev norm in (\ref{genius}) to a metric structure on $T \ca$. The results in this paper can therefore be viewed as a stepping stone towards gauge-covariant Hilbert space representation of the $\mathbf{QHD}(M)$ algebra. 
Concerning the first question then it is plausible that the Lorentz symmetry will be modified with a scale transformation. A modification of the Lorentz symmetry at the Planck scale could be within experimental bounds \cite{Jacobson}. 

The fact that the ultra-violet regularisation emerges as a part of representation theory suggest, in our opinion, that it is not a computational artefact.

\vspace{1cm}
\noindent{\bf\large Acknowledgements}\\

We would like to thank Professor Yum-Tong Siu for providing a reference on the Sobolev norm. We would also like to thank Professor Daniel Grieser for help on the Hodge-Laplace operator.
JMG would like to express his gratitude towards Ilyas Khan, United Kingdom, for his generous financial support. JMG would also like to express his gratitude towards the following list of sponsors:  Ria Blanken, Niels Peter Dahl, Simon Kitson, Rita and Hans-J\o rgen Mogensen, Tero Pulkkinen and Christopher Skak for their financial support, as well as all the backers of the 2016 Indiegogo crowdfunding campaign, that has enabled this work.

\end{document}